\documentstyle[12pt,epsf]{article}

\begin{document}
\begin{flushright}
SLAC--PUB-7609(rev 2)\\
June 1999\\  
\end{flushright}

\bigskip\bigskip
\begin{center}

{\bf\large
SOLUTION OF A RELATIVISTIC THREE BODY PROBLEM}

\bigskip

H. Pierre Noyes
\footnote{\baselineskip=12pt
Work supported by Department of Energy contract DE--AC03--76SF00515.}\\
Stanford Linear Accelerator Center\\
Stanford University, Stanford, CA 94309\\
and\\
E.D.Jones
\footnote{\baselineskip=12pt
Work supported by Department of Energy contract W-7405-Eng-48.}\\
Lawrence Livermore National Laboratory\\
Livermore, CA 94551\\
\end{center}
\vfill
\begin{abstract}
Starting from a relativistic s-wave scattering length model
for the two particle input we construct an unambiguous, unitary
solution of the relativistic three body problem given
only the masses $m_a,m_b,m_c$ and the masses of the two
body bound states $\mu_{bc},\mu_{ca},\mu_{ab}$.   
\end{abstract}
\vfill
\begin{center}
Submitted to {\it Few Body Systems: Tjon Festschrift Issue}; revised
ms. resubmitted June 1998; second revision submitted June 1999. 
\end{center}

\newpage

\section{INTRODUCTION}

Some time ago one of us made an analysis of what was then needed to
``solve'' the three nucleon problem\cite{Noyes68}. The conclusions were
rather discouraging. In particular, it appeared to us that the enormous
effort that Tjon and his collaborators, and a number of other few body
nuclear physicists, were making to obtain and use ``realistic'' two-nucleon
potential models to calculate the triton wave function and related
problems were necessarily plagued by ambiguities which it would be
difficult to remove. We eventually came to the conclusion that
a ``zero-range'' or ``on-shell'' treatment of the problem might
aid in simplifying the contact between these efforts and experiment.
Unfortunately our efforts did not lead to the desired result in
the form of a unique theoretical analysis for reasons we discuss
to some extent in our paper reporting this body of
research\cite{Noyes82}. The reasons are rather complicated, but can
be briefly summarized by the statement that {\it only} a finite
particle number {\it relativistic} scattering theory can provide the
needed framework. This paper presents a relativistic scattering length
model which we believe will be 
a practical first step toward creating such a fundamental theory.   

The key to understanding how we can create a simple, soluble 
and unitary model for a relativistic three body system is to
realize that we embed the two body subsystems in the three body space from the
start. We consider two particles with invariant masses $m_a$, $m_b$
which scatter elastically when  Mandelstam variable $s$ (square of the invariant
four momentum) lies in the range $(m_a+m_b)^2 \leq s \leq M_{th}^2$;
here $M_{th}$ is the total energy of the first inelastic (particle
creation) threshold in
the zero 3-momentum reference frame and  $k(s)$ is the magnitude of 
the momentum of either particle in this frame. 
On 3-momentum and energy shell, we write $s=M^2$ with the conventional
algebraic connections\cite{Barnettetal96} to the energies of
the free particles outside the elastic scattering volume:
\begin{eqnarray}
e_a^2-m_a^2&=&p^2=e_b^2-m_b^2\nonumber\\
M^2&=&(e_a+e_b)^2-|\vec p_a +\vec p_b|^2\nonumber\\
|\vec p|(M;m_a,m_b)&=&{[(M^2-(m_a+m_b)^2)
(M^2-(m_a-m_b)^2)]^{{1\over 2}}\over 2M}\\
(m_a+m_b)^2 \leq s&=&M^2 \leq M_{th}^2\nonumber
\end{eqnarray}
As in our non-relativistic treatment\cite{Noyes82}, we consider
only an interaction in which no angular momentum is transferred.
We also analytically continue to values of $s\neq M^2$, keeping
3-momentum conserved as in the non-relativistic case, but using
this kinematics and two-particle amplitudes embedded in a multi-particle
space. As was pointed out to one of us (HPN) by J.V.Lindesay
\cite{Lindesay97} in commenting on a preliminary version of this paper,
the correct way to carry out this analytic continuation is to take 
\begin{equation} 
k_{ab}(s;m_a+m_b) = +\sqrt{{[s-(m_a+m_b)^2][s-(m_a-m_b)^2]\over
4s}}   
\end{equation}
This has the effect of embedding our two-particle interaction in
a multi-particle space.
We were earlier led to a version of this ``off shell extension''
 by other considerations\cite{McGoveran&Noyes91}. We are much indebted
to Castillejo\cite{Castillejoc90} for pointing out to us in
a discussion of \cite{McGoveran&Noyes91} that we had, in fact, implicitly assumed
we were in a multi-particle space.

Whatever the off-shell extension, we can insure on-shell unitarity 
for the scattering amplitude
$T(s)$ with the normalization $Im \ T(s) =k(s)|T|^2$ in the
elastic scattering region by using the scattering length formula
\begin{equation}
T(s) = {1\over \gamma -ik(s)}= {e^{i\delta (s)} sin \ \delta (s)\over k(s)}
\end{equation}
where $\gamma$ is any finite constant. To generate a bound state pole
at $s=\mu_{ab}^2$ we can  (for $\mu^2_{ab} > (m_a-m_b)^2$ ) take
\begin{equation}
4\mu_{ab}^2\gamma^2 =[(m_b+m_c)^2 - \mu_{ab}^2]
[\mu^2_{ab} - (m_a-m_b)^2] > 0 
\end{equation}
and pick the branch
in the square root defining $k(s)$ by analytic continuation
below elastic scattering threshold to insure that a pole occurs
\cite{Noyes&Wong59}. As we discuss to some extent in \cite{Noyes82},
p.1869 and as Weinberg explored extensively in his quasi-particle approach to
the three body problem\cite{Weinberg63a,Weinberg63b,Weinberg64}, this prescription amounts
to assuming that the two particles composing the bound state are
``structureless''. This completes our specification of the
two particle input to our three body problem. 

In the next section we show that in our context this {\it algebraic}
construction of on-shell two body unitarity plus the algebraic form
of the Faddeev equations guarantees the three body unitarity of their
solution. This was not obvious to some people in the context of our
previous non-relativistic analysis, which led to considerable formal
complexity in the final presentation\cite{Noyes82}. In the current
context the triviality becomes manifest for the 2-2
channels below three body breakup threshold where the solution is
algebraic, as we show in Sec. 3.4. The 3 free to 3 free, coalesence and breakup
amplitudes require a more detailed analysis of the zero angular
momentum equations than we initially intended to make here. 
However, in order to meet a referee comment on an earlier version
of this paper\cite{Noyes&Jones97}, we have extended our discussion
in Sec. 3 and there provide algebraic solutions of 
the separable integral equations which follow from our assumptions
when the three free particle channels are explicitly included. 
Examples of these solutions will be explored elsewhere
\cite{Lindesayetal97}. Our concluding section contains
speculations as to where this simple result might be applied.

\section{RELATIVISTIC THREE BODY UNITARITY}
  
\subsection{Relativistic Faddeev 2-2 channel kinematics}

Our model contains three structureless particles labeled $a,b,c$,
which taken pairwise can form three bound states $(bc), (ca), (ab)$
whose only structure arises from these simple constituents. Although
we are dealing with a relativistic system, we restrict our energies
to the range which allows no particle creation; in this respect
the situation is the same as in the non-relativistic three body
problem, and the Faddeev\cite{Faddeev60,Faddeev65} channel
decomposition can be employed. Because of this simple structure,
if we start in a state with zero total 3-momentum and angular momentum,
this situation will persist so long as our interactions have no
internal degrees of freedom. This is true by hypothesis for the
model we described in the introduction. We further assume that
we start the system out with a scattering between one of the
particles (which we can pick to be $a$) and the implied bound state
pair (which will then be $bc$). Following Faddeev, we drop the
redundant index. We distinguish the bare particles from the bound states
by using the symbols $m$ and $\mu$ for their respective masses.
Then a specific scattering problem is completely specified by 
supplying numerical values for the following eight parameters:
\begin{eqnarray}
Particle \ masses &:& m_a,\ m_b, \ m_c\nonumber\\
Bound \ state \ masses &:& \mu_a, \mu_b, \mu_c\\
Invariant \ 4-momentum &:& M=\sqrt{S}\nonumber\\  
Input \ momentum &:& k_a\nonumber
\end{eqnarray}
The input momentum is the magnitude of the 3-momentum of particle $a$ 
of mass $m_a$ in the zero momentum 
frame with the bound state of particles $b$ and $c$ of mass $\mu_a$
having 3-momentum of equal magnitude
but opposite direction. We call this momentum $\kappa_a$
Calling the corresponding energy $\epsilon_a$ and the energy
of the bare particle $e_a$ {\it fixes} the
exterior kinematics as follows:
\begin{eqnarray}  
e^2_x - k^2_x &=& m^2_x ;\ \ x\in a,b,c\nonumber\\
\epsilon^2_x-\kappa^2_x &=& \mu^2_x;\ \ x\in a,b,c\\
\kappa_x &=& -k_x\nonumber\\
M_{abc} &=& e_x +\epsilon_x = M \ independent \ of \ x \nonumber
\end{eqnarray}
As can be seen from Eq.7 below, the constraint $M > m_x+\mu_x$ is
required, introducing the $\theta$-functions explicitly noted 
in Eq.52 below which mark the opening of the 2+1 channels at
their energetic thresholds; these $\theta$-functions
are required for consistency
with the orthogonality (Eq.45) and completeness (Eq.46) relations
needed in the most general case we discuss.

We will not need to leave the three body zero momentum frame
in this paper. Given $M$, we can in this restricted environment
immediately compute the input momenta because 
\begin{eqnarray}
4M^2k_x^2(\mu_x,M) = & [M^2 - (m_x + \mu_x)^2]
[M^2 - (m_x - \mu_x)^2] & = 4M^2\kappa_x^2\\
&x \ \in \ a, b, c&\nonumber
\end{eqnarray}
Thus the input momentum is not a parameter, but we still have to
specify whether the input channel is $a$ or $b$ or $c$.
Above three body breakup threshold ($M > m_a+m_b+m_c$) we must specify both the
input and the output momenta for the
``spectator'' (see Sec. 3.1). 
Because we can have one or two rearrangement 
output channels in addition to elastic scattering in the entrance
channel, the entrance channel has to be open for {\it anything}
to happen, but how many output channels will then be reached
introduces an asymmetry into the predictions.
The next problem is to insure that these various
possibilities are described by a unitary formalism.

\subsection{On Shell Faddeev Equations}

If we start in the $a$ channel, the simplest thing that can happen
is that the $bc$ pair scatter without the initial spectator $a$
being affected. These are the ``disconnected diagrams'' which initially
caused so much trouble in analyzing the three body scattering problem,
until Faddeev realized that that they could be subtracted out of
the amplitude, leaving only connected diagrams. Here we call the full
amplitude
\begin{equation}
^{(3)}T(M) \equiv \Sigma_{x,y\in a,b,c} \
^{(3)}T_{xy}(M)\theta(M-m_x-\mu_x)\theta(M-m_y-\mu_y) 
\end{equation}
where $^{(3)}T_{xy}(M)$ are the unique solutions of Eq. 10 below.

We will return to the $\theta-$functions in Sec. 3, where 
we compute only those processes allowed by the conservation laws,
except when a bound state vertex opens or closes in a single channel. 
We call the disconnected amplitude $^{(3)}T_a(M)$ and define it by
\begin{equation}
^{(3)}T_a(M) \equiv \ ^{(2)}T_{bc}([M-e_a(M)]^2)
\end{equation}
where $e_a(M)$ is to be computed from Eq. 6.
Then the relativistic version of the ``on-shell Faddeev Equations''
\cite{Noyes74} for our simple pole model become
\begin{eqnarray} 
T_{ab} - T_a\delta_{ab} &=& T_a RT_{bb} + T_a RT_{cb}
=T_a\Sigma_x \bar \delta_{ax}RT_{xb}\nonumber\\
&=&T_{aa}RT_b + T_{ac}RT_b =[\Sigma_y T_{ay}R\bar \delta_{yb}]T_b\\ 
cyclic \ &on& \ a,b,c \nonumber\\
\bar \delta_{ab}&\equiv&1-\delta_{ab}\nonumber
\end{eqnarray}
Here $R$ is the three free particle propagator, which because of our
on-shell kinematics is simply a constant whose value we will determine
from unitarity in the next two sections. The fact that the two
alternative forms of Eq. 10
define the same function is critical for time reversal invariance.
That requirement was what led us to conclude\cite{Noyes82} that our 
non-relativistic zero range model
could not be consistently defined 
when the two body amplitudes contain ``left hand cuts'' and not just
bound state or CDD\cite{CDD} poles. Algebraically the fact that
summing the multiple scattering series starting from either the first
or the last scattering --- which is what the two alternative
forms of Eq.10 express --- follows
from the fact that the two forms define the same
multiple scattering series. Convergence follows if we use two-body
input amplitudes which are always less than unity in absolute
magnitude\cite{Lindesay97}.

In our theory the convergence of the
two forms of the multiple scattering series given in Eq. 10 to the
same function suffices to prove the uniqueness of the solution.
In a normal Hamiltonian theory, one would require that the
homogeneous solutions of the two equations also must be proved to be
identical. A related question is to ask ``what operator must be
diagonalized in order to make the spectral expansion (46).''
Both questions are discussed more fully in the
non-relativistic context when taking the ``zero range limit'' 
(see Ref. 21). Note that the same dispersion
relation used to establish the unitarity of the two-body input
in the non-relativistic case also applies to the relativistic
scattering length model used here. 

There is no known unambiguous 
way to go from {\it any} field-theoretic description
of strongly interacting particles to a non-relativistic potential
model of the class used {\it phenomenologically} in nuclear
physics\cite{Noyes68}.
Ref.21 tried to remedy this defect by starting from a dispersion-
theoretic representation of the two body amplitudes which {\it is}
the non-relativistic limit of dispersion-theoretic relativistic
S-Matrix amplitudes (cf. Ref.25), leading to a ``left hand cut''
representing virtual relativistic particle exchanges (eg pions).
The difficulty which showed up in implementing this idea was
that this approach cannot be made general enough. In particular,
the Low equation (which, superficially, would seem to allow us
to construct the non-relativistic potential from knowledge
of the two-body phase shifts {\it and} the left hand cuts
in the corresponding amplitudes)
cannot be constructed. Specifically, the requirement of time
reversal invariance for the two-body potential model which results from
the attempted construction
cannot be satisfied when the dispersion-theoretic input has
left hand cuts. Consequently any bound state pole
in the two body-input will not vanish in the zero range limit,
and corresponds to a CDD pole (Ref.7) as noted above.
Consequently, some relativistic theory must be invoked to
provide the parameters of this pole. As mentioned in Sec.3.4
we use here the ``handy-dandy formula'' connecting masses to coupling
constants. We conclude that there is no direct way to relate our model to a
non-relativistic Hamiltonian model, and  hence that our completeness
relation is part our model; it need not be derived. 

The second point to emphasize is that our two-body  bound states are
{\it structureless}. They only appear in our three body system
above energetic threshold, when the third particle can play
the kinematic role of a ``spectator''. For the same reason,
the equations in this paper cannot lead directly to three body
bound states, which is why we do not need the (non-existent)
``homogeneous solutions'' to establish the equivalence of the
two orders of the operators in Eq.10. The convergence of the
summation of the two multiple scattering series to the same result
suffices. Consistent treatment of three body bound states
requires us to embed this model in a four body space in
order to provide a kinematic ``spectator'', and is not
attempted in this paper.  

\subsection{Three Particle Unitarity from Two Particle Unitarity}

That the unitarity of the {\it off-shell} Faddeev equations follows
from the unitarity of the two body input amplitudes was shown
by Freedman, Lovelace and Namyslowski\cite{FLN66} and independently
by Kowalski\cite{Kowalski70}, who taught HPN the simple algebraic
proof given below. The key is to write the unitarity condition
on the two-body amplitude $t_a$ {\it in the three body space} as
\begin{equation} 
t_a(M) - t_a^*(M) = t_a(M)(R-R^*)t_a^*(M)
\end{equation}
which determines the normalization of $R$ in terms of the normalization
of $t_a$. In order to avoid kinematic factors, it is convenient
to use 
\begin{equation}
t_a(M)=[e^{i\delta_s(s)}sin \ \delta_a(s)]_{s=[M-e_a(M)]^2} =[k_{bc}(s) \
^{(2)}T_{bc}(s)]_{s=[M-e_a(M)]^2}
\end{equation}
so that $t-t^*=2i \ tt^*$. Then $R=+i$ provides a channel
independent propagator in the three particle space.
The proper thresholds for the opening of the various channels
are provided by the $\theta-$functions in Eq. 8. 
Using the same normalization, the unitarity
condition on the three body channel amplitudes is then simply
\begin{equation}
T(M) - T^*(M) = T(M)(R-R^*)T^*(M)
\end{equation}
and below breakup threshold the on-shell Faddeev equations become 
the algebraic equations
\begin{eqnarray} 
T_{ax} - t_a\delta_{ax} &=& +it_a(T_{bx} + T_{cx})
=+it_a\Sigma_y \bar \delta_{ay}T_{yx}\nonumber\\
&=&+i((T_{ay}+T_{az})t_x =+i[\Sigma_y T_{ay}\bar \delta_{yx}]t_x 
\end{eqnarray}

The proof of unitarity follows the same steps taken in\cite{Noyes74}, Eq.'s (2)-(5),
but below breakup threshold
these are now actually algebraic rather than symbolic.
If in $T_{ab}$ we call $L_a=\Sigma_b T_{ab}$ and $F_b=\Sigma_a T_{ab}$,
we clearly have that 
\begin{equation}
T(M) =\Sigma_a L_a = \Sigma_b F_b
\end{equation}  
Then the unitarity condition we wish to prove becomes
\begin{eqnarray}
T - T^* &{?\atop =}& \nonumber\\
\Sigma_x F_x(R-R^*)L_x^*
&+& \Sigma_{x,y}\bar \delta_{xy}F_x(R-R^*)L_y^* 
\end{eqnarray}
where we have separated out the term where the indices differ
so that we can make use of the two body unitarity condition.
This is possible because, in the current notation, we can
rewrite the Faddeev equations (Eq.10) as 
\begin{eqnarray}
F_x &=& (1 - \Sigma_w F_w\bar \delta_{wx}R)t_x
\nonumber\\
L_x^* &=& t^*_x(1 - \Sigma_zR^*\bar \delta_{zx} L_z^*)
\end{eqnarray}
Consequently the equation to be proved becomes
\begin{eqnarray}
T - T^* &{?\atop =}& \nonumber\\
\Sigma_x (1 - \Sigma_w F_w\bar \delta_{wx}R)
&t_x(R-R^*)t_x^*& 
 (1 - \Sigma_zR^*\bar \delta_{zx} L_z^*) \\
&+& \Sigma_{x,y}\bar \delta_{xy}F_x(R^*-R)L_y^* \nonumber
\end{eqnarray}
We can now take the critical step of substituting $t_x - t_x^*$
for $t_x(R-R^*)t_x^*$ and find that
\begin{eqnarray}
&T - T^* {?\atop =}& \nonumber\\
&\Sigma_x (1 - \Sigma_w F_w\bar \delta_{wx}R)t_x -
t_x^* (1 - \Sigma_zR^*\bar \delta_{zx} L_z^*)& \\
-  &\Sigma_{x,y}
[t_x -\Sigma_w  F_w\bar \delta_{wx}Rt_x -F_x]R^*\bar \delta_{xy}L_y^*&\nonumber\\
+  &\Sigma_{x,y} F_x R\bar \delta_{xy}
[t_y^* - t_y^*\Sigma_zR^*\bar \delta_{zx} L_z^* -L_y^*]&\nonumber\\
&=& T - T^* \ \ \ Q.E.D. \nonumber
\end{eqnarray}
where the unwanted terms vanish because $F_x$ and $L_y^*$ are solutions
of the Faddeev equations.

\section{SOLUTION OF THE ON-SHELL FADDEEV EQUATIONS}

\subsection{3-3 Kinematics}

In  order to properly 
relate the amplitudes for elastic and rearrangement collisions 
below and above breakup threshold
it is useful
to first express the amplitudes as operators in an orthonormal and
complete space and then reduce them to integral or algebraic equations.
In our non-relativistic treatment\cite{Noyes82} this happened
automatically because there we took the ``zero range limit'' of equations
originally written in a larger space. Here we make a new approach by
formulating the operators directly in the on-shell space of empirically
observable particle and bound state momenta. For 3-3 collisions, under the restriction
to s-channel driving terms and total angular momentum zero (and ignoring
for the moment the two-body bound states and channels associated with
them), this is simply the
space of the relativistic Dalitz plot. 
We modify the standard notation\cite{Barnettetal96} to accommodate the Faddeev channel 
decomposition ($a$ the spectator of a $bc$ pair interaction or scattering) as follows.
For single free particles $p_i = (E_i,{\vec p}_i)$, $i \in a,b,c$,
and $p_i^2 =E_i^2 - {\vec p}_i\cdot {\vec p}_i=m_i^2$. 
Then, for $i,j,k$
cyclic or anti-cyclic on $a,b,c$, if $p_i$ is the four-momentum of
the spectator, we define the Mandelstam invariant $s_i$ for this
channel by
\begin{equation}
s_i \equiv m_{jk}^2 =(p_j+p_k)^2 = (E_j+E_k)^2 - 
({\vec p}_j+{\vec p}_k )\cdot ({\vec p}_j+{\vec p}_k )  
\end{equation}
In an arbitrary coordinate system
\begin{equation}
P\equiv p_a+p_b+p_c; P^2\equiv M^2 = (E_a+E_b+E_c)^2 - 
({\vec p}_a+{\vec p}_b+ {\vec p}_c)\cdot ({\vec p}_a+{\vec p}_b+ {\vec p}_c)
\end{equation}
It follows immediately that in the coordinate system where ${\vec P}
\equiv ({\vec p}_a+{\vec p}_b+ {\vec p}_c) = 0$, that
\begin{eqnarray}
s_i &=& M^2 + m_i^2 - 2ME_i\nonumber\\
s_a+s_b+s_c &=& M^2 + m_a^2 +m_b^2 +m_c^2 \equiv \Sigma^2
\end{eqnarray}
and that, cf (34.20b), with $|{\vec p}_a|^2 \rightarrow k_a^2$,
\begin{eqnarray}
4M^2k_a^2(s_a) &=& [M^2 -(s_a^{{1\over 2}}+m_a)^2]
[M^2 -(s_a^{{1\over 2}}-m_a)^2]\nonumber\\
&=& (M^2-m_a^2 -s_a)^2 -4m_a^2s_a\\
&=& (M^2 +m_a^2 -s_a)^2 -4M^2m_a^2\nonumber
\end{eqnarray}
allowing us to define $k_a(s_a) \equiv |{\vec p_a}|$ 
in this zero 3-momentum frame.

The magnitudes of the three momenta $k_a,k_b,k_c$, 
or equivalently (cf. Eq.'s 22 and 23) the three invariants 
$s_a,s_b,s_c$ , specify a rigid triangle. The orientation of this
triangle relative to some space fixed system of axes in which the
system as a whole has zero 3-momentum can be specified by three
Euler angles, $(\alpha,\beta,\gamma)$, cf. Eq. 34.18. In, for example, 
Osborn's treatment\cite{Osborn67}, the three degrees of freedom connecting
the initial to the final state are discretized as a rotation matrix
$d^J_{\lambda,\lambda'}(\theta)$ and $cos \ \theta$ is expressed
in terms of the scalar momenta $k,k'$. Under our s-channel, zero total
angular momentum assumption, this rotation matrix is simply a constant,
which is a way of seeing why we have only three degrees of
freedom even before we go on shell (restrict ourselves to the interior
of the Dalitz plot by fixing $M$ in Eq.22).

\subsection{Separable integral equations for 3 free- 3 free
scattering}

Define the operators ${\cal M}_{ax}\equiv {\cal T}_a\delta_{ax} +{\cal W}_{ax}$,
making the Faddeev equations starting from entrance channel $a$ read
\begin{eqnarray}
{\cal W}_{aa} -{\cal T}_a{\cal R}({\cal W}_{ba} + {\cal W}_{ca})
&=& 0 \nonumber\\
- {\cal T}_b{\cal R}{\cal W}_{aa} + {\cal W}_{ba} 
- {\cal T}_b{\cal R}{\cal W}_{ca}
&=& {\cal T}_b{\cal R}{\cal T}_a\\
- {\cal T}_c{\cal R}({\cal W}_{aa} + {\cal W}_{ba}) + {\cal W}_{ca}
&=& {\cal T}_c{\cal R}{\cal T}_a
\nonumber
\end{eqnarray}
Separate ${\cal W}_{ax}$ into the four amplitude operators
\begin{eqnarray}
{\cal W}_{ax}&=& {\cal A}_{ax} + {\cal B}_{ax}
+ {\cal C}_{ax} + {\cal D}_{ax}\\
{\cal A}_{ax} &:& Anelastic \ scattering \nonumber\\
{\cal B}_{ax} &:& Breakup \nonumber\\
{\cal C}_{ax} &:& Coalesence \nonumber\\
{\cal D}_{ax} &:& 3 \ free - 3 \ free \ scattering \nonumber
\end{eqnarray}
Assume that the first three are zero. Then the operator Faddeev
equations for the 3-free to 3-free amplitude  
starting in entrance channel $a$ are
\begin{eqnarray}
{\cal D}_{aa} -{\cal T}_a{\cal R}({\cal D}_{ba} + {\cal D}_{ca})
&=& 0 \nonumber\\
- {\cal T}_b{\cal R}{\cal D}_{aa} + {\cal D}_{ba} - {\cal T}_b{\cal R}{\cal D}_{ca}
&=& {\cal T}_b{\cal R}{\cal T}_a\\
- {\cal T}_c{\cal R}({\cal D}_{aa} + {\cal D}_{ba}) + {\cal D}_{ca}
&=& {\cal T}_c{\cal R}{\cal T}_a
\nonumber
\end{eqnarray}
and the two obvious cyclic permutations for entrance channels $b$ and
$c$. The alternative form, with exit channel (rather than entrance channel)
specified as $a$ is 
\begin{eqnarray}
{\cal D}_{aa} -({\cal D}_{ab} + {\cal D}_{ac}){\cal R}{\cal T}_a
&=& 0 \nonumber\\
- {\cal D}_{aa}{\cal R}{\cal T}_b + {\cal D}_{ab} - {\cal D}_{ac}{\cal R}{\cal T}_b
&=& {\cal T}_a{\cal R}{\cal T}_b\\
- ({\cal D}_{aa} + {\cal D}_{ab}){\cal R}{\cal T}_c + {\cal D}_{ac}
&=& {\cal T}_a{\cal R}{\cal T}_c
\nonumber
\end{eqnarray}
 
To convert these into one-variable integral equations, we assume the
orthogonality and completeness relations
\begin{eqnarray}
<s_x|s_y'> &=& \delta_{xy}\delta(s_x-s_y')\\
\int_{(m_y+m_z)^2}^{(M-m_x)^2}ds_x |s_x><s_x| 
&=&{\bf 1}\nonumber
\end{eqnarray}
for the states onto which we project. Here, and 
from now on, the kinematic constraint
(cf. Eq.22) on the variables
\begin{equation}
s_a+s_b+s_c = M^2 + m_a^2 +m_b^2 +m_c^2 \equiv \Sigma^2
=s_a'+s_b'+s_c'
\end{equation}
is to be understood.
The matrix elements of the operators are taken to be
\begin{eqnarray}
<s_x|{\cal R}|s_y'> &=& R\bar \delta_{xy}\theta(M-m_a-m_b-m_c)
\nonumber\\
<s_x|{\cal T}_x|s_y'> &=& \delta_{xy}\delta(s_x-s_y')e^{i\delta_x}sin \
\delta_x(s_x,M)\theta(s_x- (m_y+m_z)^2)\\
<s_x|{\cal D}_{xy}|s_y'> &=& D_{xy}(s_x;M;s_y')\theta(M-m_a-m_b-m_c)\nonumber
\end{eqnarray}
where $R$ is a constant to be fixed by requiring on-shell unitarity
and $\bar \delta_{xy}= 1 - \delta_{xy}$.
It is convenient to also define $t_x(s_x,M) \equiv e^{i\delta_x}sin \
\delta_x(s_x,M)$. Then, by taking matrix elements of the operator
equations between the appropriate bras and kets, and invoking the
completeness relation to bring in the appropriate integrals, we find
that, for $a$ the entrance channel, 
\begin{eqnarray}
&D_{aa}(s_a;s_a')
-t_a(s_a)R[\int_{(m_a+m_c)2}^{(M-m_b)^2}ds_b''D_{ba}(s_b'';s_a') 
+ \int_{(m_a+m_b)2}^{(M-m_c)^2}ds_c''D_{ca}(s_c'';s_a')]&\nonumber\\
&=0& \nonumber\\
&- t_b(s_b)R\int_{(m_b+m_c)2}^{(M-m_a)^2}ds_a''{\cal D}_{aa}(s_a'';s_a') 
+ D_{ba}(s_b;s_a') 
-t_b(s_b)R\int_{(m_a+m_b)2}^{(M-m_c)^2}ds_c''D_{ca}(s_c'';s_a')&\nonumber\\
&= t_b(s_b)Rt_a(s_a')&\\
&- t_c(s_c)R[\int_{(m_b+m_c)2}^{(M-m_a)^2}ds_a''D_{aa}(s_a'';s_a') 
+ \int_{(m_a+m_c)2}^{(M-m_b)^2}ds_b''D_{ba}(s_b'';s_a')] 
+ {\cal D}_{ca}(s_c;s_a')&\nonumber\\
&= t_c(s_c)Rt_a(s_a')&\nonumber
\end{eqnarray}
Here we have suppressed the fixed argument $M$ in the functions and
retained only the variables $s_x, s_x',s_x''$ of the integral equation and
the fixed parameter $s_a'$ describing the initial state.
The alternative form of the equations (Eq. 27) reverses the role of variables
and parameters, leading to
\begin{eqnarray}
&D_{aa}(s_a';s_a)
-[\int_{(m_a+m_c)2}^{(M-m_b)^2}ds_b''D_{ab}(s_a';s_b'') 
+ \int_{(m_a+m_b)2}^{(M-m_c)^2}ds_c''D_{ac}(s_a';s_c'')]Rt_a(s_a)
&\nonumber\\
&=0& \nonumber\\
&- \int_{(m_b+m_c)2}^{(M-m_a)^2}ds_a''{\cal D}_{aa}(s_a';s_a'')Rt_b(s_b) 
+ D_{ab}(s_a';s_b) 
-\int_{(m_a+m_b)2}^{(M-m_c)^2}ds_c''D_{ac}(s_a';s_c'')Rt_b(s_b)
&\nonumber\\
&= t_a(s_a')Rt_b(s_b)&\\
&- [\int_{(m_b+m_c)2}^{(M-m_a)^2}ds_a''D_{aa}(s_a';s_a'') 
+ \int_{(m_a+m_c)2}^{(M-m_b)^2}ds_b''D_{ba}(s_a';s_b'')]Rt_c(s_c) 
+ {\cal D}_{ca}(s_c;s_a')&\nonumber\\
&= t_a(s_a')Rt_c(s_c)&\nonumber
\end{eqnarray}
We see that Eq.31 and Eq.32 are, as the sub-section title claims, separable integral
equations for one-variable, one-parameter functions. But
the two alternative forms have to be solved for all nine functions
and not just for three of them before we can even ask the question
as to whether they both define the {\it same} nine functions
$D_{xy}(s_x;M;s_x)$. Until we have made this proof, it will be safer
to call the solutions of Eq. 31 (and their two cyclic permutations)
$L_{xy}(s_x;M;s_y')$ with the understanding that $s_x$ is the variable
and $s_y'$ the parameter while the solutions of Eq. 32 will be
designated by $R_{xy}(s_x';M;s_y)$.

\subsection{Solution of the 3-3 problem}

To solve these equations we first define
\begin{eqnarray}
\bar t_x &\equiv& \int_{(m_y+m_z)^2}^{(M-m_x)^2}ds_x
t_x(s_x;M)\\
l_{xy}(s_y)&\equiv& \int_{(m_y+m_z)^2}^{(M-m_x)^2}ds_x 
L_{xy}(s_x;M;s_y)\nonumber\\
r_{xy}(s_x)&\equiv& \int_{(m_z+m_x)^2}^{(M-m_y)^2}ds_y 
R_{xy}(s_x;M;s_y)\nonumber
\end{eqnarray}
Then, integrating Eq. 31 over the variables yields
\begin{equation}
\left(
\begin{array}{ccc}
 1 & -\bar t_a R & -\bar t_a R\\[1ex]  
-\bar t_b R & 1 & -\bar t_b R\\
-\bar t_c R & -\bar t_c R & 1 \end{array}\right)
\left(\begin{array}{c}
l_{aa}(s_a')\\
l_{ba}(s_a')\\
l_{ca}(s_a')\end{array}\right)=
\left(\begin{array}{c}
0\\
\bar t_bRt_a(s_a')\\
\bar t_cRt_a(s_a')\end{array}\right)
\end{equation}
While integrating Eq. 32 over the variables yields
\begin{equation}
\left(
\begin{array}{ccc}
 1 & -\bar t_a R & -\bar t_a R\\[1ex]  
-\bar t_b R & 1 & -\bar t_b R\\
-\bar t_c R & -\bar t_c R & 1 \end{array}\right)
\left(\begin{array}{c}
r_{aa}(s_a')\\
r_{ab}(s_a')\\
r_{ac}(s_a')\end{array}\right)=
\left(\begin{array}{c}
0\\
t_a(s_a') R\bar t_b\\
t_a(s_a') R\bar t_c\end{array}\right)
\end{equation}

These equations are clearly algebraic and, together with
the two cyclic permutations, define the functions
$l_{xy}(s_y')$ and $r_{xy}(s_x')$ for any value of the entrance
or exit parameters allowed by the fixed value of $M$.
Assuming that Eq. 34 has been inverted (see below), we then can
insert the solution in Eq. 31 to specify the full solution
\begin{eqnarray}
L_{aa}(s_a;M;s_a')&=& t_a(s_a;M)R[l_{ba}(s_a')+ l_{ca}(s_a')]\nonumber\\
L_{ba}(s_b;M;s_a')&=& t_b(s_b;M)R[t_a(s_a';M)+
l_{aa}(s_a')+ l_{ca}(s_a')]\\
L_{ca}(s_c;M;s_a')&=& t_c(s_c;M)R[t_a(s_a';M) +
l_{aa}(s_a')+ l_{ba}(s_a')]\nonumber
\end{eqnarray}
while if Eq. 35 has been inverted we find that
\begin{eqnarray}
R_{aa}(s_a';M;s_a)&=& [r_{ab}(s_a')+ r_{ac}(s_a')]Rt_a(s_a;M)\nonumber\\
R_{ab}(s_a';M;s_b)&=& [t_a(s_a';M)+
r_{aa}(s_a')+ r_{ac}(s_a')]Rt_b(s_b;M)\\
R_{ac}(s_a';M;s_c)&=& [t_a(s_a';M)+
r_{aa}(s_a')+ r_{ab}(s_a')]Rt_c(s_c;M)R\nonumber
\end{eqnarray}
It still remains to prove that, when we complete the system by
supplying the remaining permutations, the two {\it different}
routes by which we arrive at the solutions define the {\it same}
nine functions. 

One way to do this is to define
\begin{eqnarray}
t_{xy}&\equiv&\bar \delta_{xy}\bar t_x R \bar t_y = t_{yx}\\ 
l_{xy}&\equiv& \int_{(m_y+m_z)^2}^{(M-m_x)^2}ds_x 
\int_{(m_z+m_x)^2}^{(M-m_y)^2}ds_y L_{xy}(s_x;M;s_y)\nonumber\\
r_{xy}&\equiv& \int_{(m_y+m_z)^2}^{(M-m_x)^2}ds_x 
\int_{(m_z+m_x)^2}^{(M-m_y)^2}ds_y R_{xy}(s_x;M;s_y)\nonumber
\end{eqnarray}

Then, integrating Eq. 34 over the remaining parameter yields
\begin{equation}
\left(
\begin{array}{ccc}
 1 & -\bar t_a R & -\bar t_a R\\[1ex]  
-\bar t_b R & 1 & -\bar t_b R\\
-\bar t_c R & -\bar t_c R & 1 \end{array}\right)
\left(\begin{array}{c}
l_{aa}\\
l_{ba}\\
l_{ca}\end{array}\right)=
\left(\begin{array}{c}
0\\
t_{ba} (=t_{ab})\\
t_{ca} (=t_{ac})\end{array}\right)
\end{equation}
while integrating Eq. 35  over the remaining parameter yields
\begin{equation}
\left(
\begin{array}{ccc}
 1 & -\bar t_a R & -\bar t_a R\\[1ex]  
-\bar t_b R & 1 & -\bar t_b R\\
-\bar t_c R & -\bar t_c R & 1 \end{array}\right)
\left(\begin{array}{c}
r_{aa}\\
r_{ab}\\
r_{ac}\end{array}\right)=
\left(\begin{array}{c}
0\\
t_{ab} (=t_{ba})\\
t_{ac} (=t_{ab})\end{array}\right)
\end{equation}
We see immediately that the two alternative forms define the same
algebraic matrix and hence that $l_{xy}=r_{yx}$. Because of the
symmetry of the driving terms $l_{xy}=l_{yx}$, $r_{xy}=r_{yx}$ 
and we are free to define
\begin{equation}
z_{xy} \equiv l_{xy} = r_{xy}=l_{yx} = r_{yx}
\end{equation}
The inversion is algebraic and straightforward\cite{HANS}. Explicitly,
with
\begin{equation} 
R=i 
\end{equation}
we find that 

\begin{equation}
\left(\begin{array}{c}
z_{aa}\\
z_{ba}\\
z_{ca}\end{array}\right)=
\left(\begin{array}{c}
-\bar t_a \bar t_b -\bar t_a \bar t_c -2i \bar t_a \bar t_b \bar t_c\\

i\bar t_b  - \bar t_b \bar t_c\\
i\bar t_c  - \bar t_c \bar t_b\end{array}\right)
{\bar t_a\over 1+\bar t_a \bar t_b +\bar t_b \bar t_c +\bar t_c \bar
t_a  +2i  \bar t_a \bar t_b \bar t_c} 
\end{equation}

We have now solved the 3-3 problem (with no bound states) because 
we can reconstruct the variable and parameter content of the solution as
\begin{eqnarray}
& For \ {\cal A}={\cal B}={\cal C} =0 &\nonumber\\
M_{aa}(s_a;M;s_a')&=&t_a(s_a,M)\delta(s_a-s_a') 
+ t_a(s_a,M)z_{aa}t_a(s_a',M) \nonumber\\
M_{ba}(s_b;M;s_a')&=& t_b(s_b,M)z_{ba}t_a(s_a',M)\\
M_{ca}(s_c;M;s_a')&=&
t_c(s_c,M)z_{ca}t_a(s_a',M)\nonumber
\end{eqnarray}
Clearly, the remaining six amplitudes can be written in the same way.

The Kowalski version of the unitarity proof is now algebraic and
trivial, provided we use the state normalization which makes
two-particle unitarity require that $t-t^*=2itt^*$ and
$R-R^*=2iRR^*$, which forces us to choose the constant $R=i$.

\subsection{Coalesence, Breakup, Anelastic Scattering coupled to 3 free-3
free Scattering}

If we now include $N_a +N_b +N_c$ two-body bound states $|\mu_x^{n_x}>$ 
with $x\in a,b,c$ and $n_x\in 1,2,..,N_x$ the orthonormality condition
must be extended to include
\begin{eqnarray}
<\mu_x^{n_x}|\mu_{x'}^{n'_{x'}}>
&=&\delta_{xx'}\delta_{n_xn'_{x'}}\\
<s_x|\mu_x^n> &=& 0 \nonumber
\end{eqnarray}
and the completeness equation becomes
\begin{equation}
\int_{(m_y+m_z)^2}^{(M-m_x)^2}ds_x
|s_x><s_x| +\Sigma_{n_x=1}^{N_x}
|\mu_x^{n_x}><\mu_x^{n_x}| ={\bf 1}
\end{equation}

Although the propagator ${\bf \cal R}\bar \delta_{xy} $ remains constant,
using our normalizations in this extended space, we can add a new
dynamical element to the system if we allow (as we can, conserving
on-shell energy and 3-momentum) the 2-2 scattering operators 
${\cal T}_a(M)$,
whose matrix elements in the 3-3 part of the space are given above,
to couple 1,2 states (anelastic scattering) to 3 free particle states 
(breakup) or visa versa (coalesence)  
via the three free particle propagator ${\cal R}$
preserving probability conservation (unitarity). In
constructing our specific way of providing this coupling, we have been
guided by (a) the success of the dispersion-theoretic non-relativistic approximation 
of single pion exchange in predicting the $^1S_0$ shape parameter in
the archetypal strong interaction problem (nucleon-nucleon scattering
\cite{Noyes&Lipinski71,Noyes72}),
(b) Faddeev's analysis of the role of the ``essential singularities'' in the
non-relativistic case and (c) the success of the ``handy dandy
formula'' connecting masses to coupling
constants\cite{McGoveran&Noyes91,Noyes97}. 

We note first that the thresholds for the
opening of the elastic and anelastic channels can be uniquely
specified by the thresholds:
\begin{equation}
M_{n_x}\equiv m_x + \mu_x^{n_x}
\end{equation}   
and (assuming no degeneracies and at least one bound state) uniquely 
ordered by some integer parameter $1\leq n \leq N_a+N_b+N_c$.
So we can always specify and order uniquely the parameters 
\begin{equation}
M_{n_x}^1 < M_{n_{x'}}^n< M_{n_{x''}}^{n+1} < m_a+m_b+m_c; 
\ \ \ x,x',x'' \in a,b,c;
\ \ \ n \leq N_a+N_b+N_c
\end{equation}

Consider first only anelastic scattering ($M_{n_x}^1 < M <
m_a+m_b+m_c$). Then, relying on Faddeev's insight (b), we will
assume that the scattering operator has the matrix element
\begin{eqnarray}
&[{lim\atop s_a\rightarrow (\mu_x^{n_x})^2}]((s_a - (\mu_x^{n_x})^2)
<\mu_x^{n_x}|{\cal T}_x|\mu_y^{n_y'}>) =&\\
&\delta_{xy}\delta_{n_xn_x'}
\ i\Gamma_x^{n_x}(k_x^{n_x},M)\theta(M-m_x-\mu_x^{n_x})&\nonumber 
\end{eqnarray} 
Here $\Gamma_x^{n_x}(k_x^{n_x},M)$ is the residue at the pole of the two-body
scattering amplitude at $s_x=(\mu_x^{n_x})^2$ and for our scattering 
length model is given by (cf.Eq.'s 2,3,4,8)
\begin{equation}
\Gamma_x^{n_x}(k_x^{n_x},M)={\gamma_x^{n_x}\over M}
\sqrt{[M^2-(m_x+\mu_x^{n_x})^2][M^2-(m_x-\mu_x^{n_x})^2]}
\end{equation}
This is a simple algebraic consequence of the algebraic structure of
our two particle input model; The $\theta$-function in Eq.49 makes
this limit consistent with the orthogonality (45) and completeness
(46) relations.
Both below and above breakup threshold we model the opening and
closing of the bound state vertex by assumung that
\begin{eqnarray}
<\mu_x^{n_x}|{\cal R}|\mu_y^{n_y'}> &=&\bar \delta_{xy}R\nonumber\\ 
<\mu_x^{n_x}|{\cal R}|s_y> &=&\bar \delta_{xy}R\\ 
<s_x|{\cal R}|\mu_y^{n_y'}> &=&\bar \delta_{xy}R\nonumber\\ 
<s_x|{\cal R}|s_y> &=&\bar \delta_{xy}R\nonumber 
\end{eqnarray}

For $M<m_a+m_b+m_c$, taking matrix elements of the Faddeev equations
between the bound states and inserting the completeness relations
as we did in the continuum case, we clearly obtain
N algebraic equations for N amplitudes where N is the
number of open channels allowed by the $\theta$-fuctions using
the fixed value of $M$. For one bound state in each channel,
we obtain a triple of three equations in three unknowns, whose solution
is of the same algebraic form as those for the $z_{xy}$ exhibited
above, with $\bar t_x$ replaced by $\Gamma_x^1$. These were the
equations we had in mind in the first version of this
paper\cite{Noyes&Jones97}; the intent of the original paper
was to explore breakup and related
problems in subsequent work\cite{Lindesayetal97}. As already noted, the
more complete treatment here was needed to meet referee comments,
for which we are grateful.

The postulate of a ``channel independent''
(indeed constant) ``three free particle propagator'' below breakup
threshold is our way of taking over the Faddeev idea that ``once
a scattering occurs, we must allow one of the pair to interact with a
third particle before anything more can happen''  into our 
on-shell theory. Another way to put it is that ``once a bound state
vertex opens up, another degree of freedom must intervene before
that pair can interact again.'' This is familiar in other Faddeev
contexts; only this particular articulation of the idea is novel.
This analysis also shows us how 
the coalesence and breakup amplitudes couple into the system (cf. Eq.'s
49,50,51).

It is now straightforward to write down the full coupled equations for 
${\cal A},{\cal B},{\cal C},{\cal D}$ above breakup threshold for any
arbitrary finite number of bound states and reduce them to algebraic
equations for the constants $z_{xy}^{n_xn_y'}$. Once solved, we can
reconstruct the full variable content of the amplitudes by writing
\begin{eqnarray}
{\cal A}_{xy}^{n_xn_y'}(M)&=&
\Gamma_x^{n_x}z_{xy}^{n_xn_y'}\Gamma_y^{n_y'}\nonumber\\
\ & \ &\theta(M-(m_x + \mu_x^{n_x}))\theta(M-(m_y + \mu_y^{n_y'}))\\
{\cal B}_{xy}^{n_y'}(s_x;M)&=&t_x(s_x,M)
z_{xy}^{n_y'}\Gamma_y^{n_y'}\nonumber\\
\ & \ &\theta(s_x-(m_a + m_b + m_c)^2)\\
{\cal C}_{xy}^{n_x}(M;s_y')&=&
\Gamma_x^{n_x}z_{xy}^{n_x}t_y(s_y',M)\nonumber\\
\ & \ &\theta(s_y'-(m_a + m_b + m_c)^2)\\
{\cal D}_{xy}(s_x;M;s_y')&=&
t_x(s_x,M)z_{xy}t_y(s_y',M)\nonumber\\
\ & \ &\theta(s_x-(m_a + m_b + m_c)^2)\theta(s_y'-(m_a + m_b + m_c)^2)
\end{eqnarray} 
Knowing $z_{xy}^{n_xn_y'}$, $z_{xy}^{n_y'}$, $z_{xy}^{n_x}$ and 
$z_{xy}$ we can now provide predictions for all
elastic, rearrangement, three body breakup and 3-3 scattering cross
sections for all energy-momentum-$J=0$ conserving processes over the
entire kinematic region $M < M_{th}$.

The two body model used assumes that we know {\it either} the binding
energy of a (single) two body bound state {\it or} the scattering
length in the isolated two-particle system {\it but not both}.
Our treatment up to this point has
implicitly assumed that our six parameters $m_x, \mu_x$, 
$x \in a,b,c$, are all consistent with known two body data
over the range of interest.
The failure of this connection, either by the two parameters
(which can be determined by different types of experiment)
being inconsistent with it, or by the departure of the predicted
elastic scattering cross section in the physical region from
experiment suffices to show that we must enrich the parameter
content of the model. This would be analagous to what happened
historically\cite{Bethe&Bacher36} in the study of neutron-proton
scattering and established the spin-dependence of nuclear forces. 
Precise experiments analyzed using related ideas have
even demonstrated the existence of pions using 
nucleon-nucleon s-wave experiments 
below 10 Mev\cite{Noyes&Lipinski71,Noyes72}. 

When it comes to the unique predictions of our three body model,
the comparison with experiment becomes richer than for the
two-body scattering length model. In particular
we can now predict {\it three} scattering lengths or their
equivalent and compare these with low energy s-wave experiments.
If any one of these fails to agree with experiment, one place
to look for an explanation is to postulate a single three-body
bound state. If our matrix formulation of the Faddeev equations,
when analytically continued below the threshold for the lowest
kinematically allowed 2-2 elastic scattering, possesses a homogeneous
solution  at some value of $M$, this value of $M$ predicts the
existence of a 3-body bound state at that invariant energy.
As already noted, this analytic continuation in our theory is only
possible after we embedded our model in an appropriate 4-body space.
If the state with the right quantum numbers is found, but the value of
$M$ differs experimentally from that predicted, we can extend our model
phenomenologically by explicitly introducing an $(abc) \leftrightarrow
(abc)$ channel in addition to the three channels we started with. 
Then we must solve four 
equations for four unknowns, but no new conceptual problems arise. 
Since direct scattering
of three free particles to three free particles is usually
too difficult to measure, particularly at three body breakup
threshold, we do not count this as a new piece of available experimental
information. However, we now have one parameter
to explain (if the bound state exists at a known mass $m_{abc}$)
four experimental numbers, giving a stringent consistency
check even using threshold data. We intend to explore consistency
conditions on these parameters elsewhere.

As a significant example, particularly relevant to Tjon's {\it oevre},
the $n,n$ and $n,p$ singlet scattering lengths and the deuteron binding energy
can be used as the three 2-2 channel parameters for the $n,n,p$
system. Only two scattering lengths ($n,d$ doublet and  
$n,d$ quartet) are measured in addition to the binding energy of the
triton. $^4a_{nd}$ is well predicted while $^2a_{nd}$ and $\epsilon_t$
are highly correlated (``Phillips line'') for reasons that can be
understood from a dispersion-theoretic point of
view\cite{Barton&Phillips69}. Of course Barton and Phillips'
explanation is consistent
with the physics underlying our on-shell approach,
so we expect our model to achieve a comparable result. In historical
fact, it was their work which helped us start thinking about the
usefulness of a more general approach in the first place.

It should be obvious from the treatment of the four body problem in our
non-relativistic paper\cite{Noyes82} that the current approach
can be readily generalized to relativistic four particle systems,
as we intend to do on another occasion. 

\section{SPECULATIONS AND CONCLUSIONS}

The example of the applicability of this model to
nuclear physics given in the last section hardly begins to
suggest the range of problems which
we believe can profitably be explored using the approach presented
here. For instance, the equivalent of the relativistic scattering
length formula used here was first written down by Bohr in 1915
\cite{Bohr15,McGoveran&Noyes91}. Viewed from the current point of view,
this makes the hydrogen atom a relativistic bound state of a proton
and an electron. This suggests looking at the three  body systems
$e,e,p$ ($H^-$), $e,p,p$ ($H_2^+$) and similar atomic systems
using the explicit model presented in this paper.

For strong interactions, we suggest treating the deuteron as a neutron-proton-pion
bound state. If we include crossing and the
Fermi-Yang model for the pion\cite{Fermi&Yang49} as a bound state of a nucleon and an
anti-nucleon the usefulness of the approach for deeply bound states
should become manifest. We suspect that relativistic models of quarks 
and quark confinement could also be attacked using these methods.  

The relativistic scattering length model employed 
here actually arose in a study of the fine structure spectrum of hydrogen
\cite{McGoveran&Noyes91}. In conjunction with combinatorial arguments,
this model leads to the Sommerfeld formula {\it without} any specific
use of the concept of ``spin'', and to an improvement of the
lowest order combinatorial calculation of the low energy fine
structure constant ($\alpha_e(m_e^2)^{-1} = 137$) by four significant
figures (to 137.03596...). Similar improvement of our understanding of
``bit-string-physics''\cite{Noyes97} can be anticipated when we
make use of the three and four body dynamics adumbered here.
We hope that others may be induced to try these simple methods
and see how far they might lead.

Because of the occasion to which this issue of Few Body Physics is
dedicated, one of us (HPN) thinks it appropriate to reiterate here,
as was stated long ago\cite{Noyes70}, that the reduction of the
three body problem from three to two continuous variables presented
by Osborn and Noyes\cite{Osborn&Noyes} was first, independently,
developed by Ahmadzadeh and Tjon\cite{Ahmadzadeh&Tjon65}.
This is yet another reminder, of which there will be many in this
issue, of how continuously useful and important Prof. Tjon's dedication
to our field has been.


\begin{thebibliography}{99}

\bibitem{Ahmadzadeh&Tjon65}  
A.Ahmadzadeh and J.A.Tjon, {\it Phys.Rev.}, {\bf 139}, B1085 (1965).   

\bibitem{Barnettetal96}
R.M.Barnett, et. al. {\it Phys.Rev.} {\bf D}, No. 1, Part I, p.175
(1996); when needed we use the equation and figure numbers as given
there..

\bibitem{Barton&Phillips69}
G.Barton and A.C.Phillips, {\it Nucl.Phys.}, {\bf A 132}, 97 (1969).

\bibitem{Bethe&Bacher36}
H.A.Bethe and R.F.Bacher, {\it Rev. Mod. Phys.}, {\bf 8}, 88-229
(1936).

\bibitem{Bohr15}
N.Bohr, {\it Phil.Mag.}, 332-335 (February 1915).

\bibitem{Castillejoc90}
L.Castillejo, private comment to HPN c.1990. 

\bibitem{CDD}
L.Castillejo, R.H.Dalitz and F.J.Dyson, {\it Phys. Rev.} {\bf 101}, 453
(1956). 

\bibitem{Faddeev60}
L.D.Faddeev, {\it JETP}, {\bf 39}, 1459 (1960).

\bibitem{Faddeev65}
L.D.Faddeev, {\it Mathematical Aspects of the Quantum Mechanical
Three-body Problem}, Davey, New York, 1965.

\bibitem{Fermi&Yang49}
E.Fermi and C.N.Yang, {\it Phys.Rev.}, {\bf 76}, 1739 (1949).

\bibitem{FLN66}
D.Z.Freedman, C.Lovelace and J.M.Namyslowski, {\it Nuovo Cimento} {\bf
43 A}, 258 (1966).

\bibitem{HANS}
We are indebted to J.C.van den Berg for performing and checking the
inversion in our context, August 1997.

\bibitem{Kowalski70}
K.L.Kowalski, private communication to HPN c. 1972.

\bibitem{Lindesay97}
J.V.Lindesay, private comment to HPN, August 7, 1997.

\bibitem{Lindesayetal97}
J.V.Lindesay, H.P.Noyes and E.D.Jones (in preparation).

\bibitem{McGoveran&Noyes91}
D.O.McGoveran and H.P.Noyes, {\it Physics Essays}, {\bf 4}, 115-120
(1991).

\bibitem{Noyes68}
H.P.Noyes, {\it Prog. Nuc. Phys.}, {\bf 10}, 355-380 (1968).

\bibitem{Noyes70}
H.P.Noyes, {\it Phys. Rev. Letters}, {\bf 25}, 324 (1970).

\bibitem{Noyes72}
H.P.Noyes, {\it Ann.Rev.Nuc.Sci.}, {\bf 22}, 465-484 (1972).

\bibitem{Noyes74}
H.P.Noyes, {\it Czech.J.Phys.}, {\bf B24}, 1205-1214 (1974).

\bibitem{Noyes82}
H.P.Noyes, {\it Phys.Rev.}, {\bf C 26}, 1858-1877 (1982).

\bibitem{Noyes97}
H.P.Noyes, ``A Short Introduction to Bit-String Physics'',
SLAC-PUB-7205, August 1997 and in {\it Merologies: Proc.ANPA 18},
T.Etter, ed., published by ANPA, c/o C.W.Kilmister, Red Tiles Cottage,
High Street, Barcombe, Lewes BN8 5DH, United Kingdom, 1997, pp 21-61.

\bibitem{Noyes&Jones97}
H.P.Noyes and E.D.Jones, ``Solution of a Relativistic Three Body
Problem'', SLAC-PUB-7609 (October 1997).

\bibitem{Noyes&Lipinski71}
H.P.Noyes and H.M.Lipinski, {\it Phys. Rev.} {\bf C 4}, 995-1002 (1971).

\bibitem{Noyes&Wong59}
H.P.Noyes and D.Y.Wong, {\it Phys.Rev.Letters}, {\bf 3}, 191(1959).

\bibitem{Omnes64}
R.L.Omnes, {\it Phys.Rev.} {\bf 134}, B1358 (1964).

\bibitem{Osborn67}
T.A.Osborn, PhD Thesis, Stanford and SLAC Report No. 79, December 1967.
See also \cite{Ahmadzadeh&Tjon65}, \cite{Omnes64}, \cite{Osborn&Noyes}.

\bibitem{Osborn&Noyes}
T.A.Osborn and H.P.Noyes, {\it Phys.Rev.Letters}, {\bf 17}, 215 (1966).

\bibitem{Weinberg63a}
S.Weinberg, {\it Phys.Rev.}, {\bf 130}, 776 (1963).

\bibitem{Weinberg63b}
S.Weinberg, {\it Phys.Rev.}, {\bf 131}, 440 (1963).

\bibitem{Weinberg64}
S.Weinberg, {\it Phys.Rev.}, {\bf 133 B}, 232 (1964).

\end{thebibliography}
\end{document}